\documentclass[prb]{revtex4-1} 

\usepackage{amsmath}  
\usepackage{amsfonts} 
\usepackage{graphicx} 

\begin{document}

\title{d'Alembert Digitized: A Wave Pulse Method for Visualizing Electromagnetic Waves in Matter and for Deriving the Finite Difference Time Domain Method for Numerically Solving Maxwell's Equations}
\author{Ross Hyman}
\email{rosshyman@att.net}
\affiliation{Materials Research Science and Engineering Center, Northwestern University, 2145 Sheridan Road, Evanston, IL 60208-3118}
\author{Nathaniel Stern}
\email{n-stern@northwestern.edu}
\affiliation{Department of Physics and Astronomy,
Northwestern University, 2145 Sheridan Rd., Evanston, IL 60208-3118}
\author{Allen Taflove}
\email{Taflove@ece.northwestern.edu}
\affiliation{Department of Electrical and Computer Engineering, Northwestern University, 2145 Sheridan Rd., Evanston, IL 60208-3118}

\date{\today}
\begin{abstract}
An alternative way of visualizing electromagnetic waves in matter and of deriving the Finite Difference Time Domain method (FDTD) for simulating Maxwell's equations for one dimensional systems is presented. The method uses d'Alembert's splitting of waves into forward and backward pulses of arbitrary shape and allows for grid spacing and material properties that vary with position. Constant velocity of waves in dispersionless dielectric materials, partial reflection and transmission at boundaries between materials with different indices of refraction, and partial reflection, transmission, and attenuation through conducting materials are derived without recourse to exponential functions, trigonometric functions, or complex numbers. Placing d'Alembert's method on a grid is shown to be equivalent to FDTD and allows for a simple and visual proof that FDTD is exact for dielectrics when the ratio of the spatial and temporal grid spacing is the wave speed, a straightforward way to incorporate reflectionless boundary conditions, and a derivation that FDTD retains second order accuracy when the grid spacing varies with position and the material parameters make sudden jumps across layer boundaries. 
\end{abstract}

\maketitle

\section{Introduction}
\subsection{Background and Motivation}
We apply a discretization of d'Alembert's splitting of waves into forward and backward pulses\cite{d'Alembert1, d'Alembert2, d'Alembert3, Wheeler} of arbitrary shape to present a visual way of deriving the physics of electromagnetic waves in matter and the Finite Difference Time Domain (FDTD) method\cite{Yee, Taflove, Sipos, Visscher, Gould, Kunz} for computationally solving Maxwell's equations in one-dimensional media. The FDTD method is a very efficient and popular way to computationally solve Maxwell's equations for the dynamics of electromagnetic waves in matter. Because it is a real time method, it is efficient for simulations requiring a broad range of frequencies such as femtosecond pulses\cite{tightpulse} and atomic systems treated semi-classically.\cite{atom} Despite its utility and popularity, the method is not covered in the standard physics Electricity and Magnetism (E\&M) text books.\cite{Jackson, Griffiths} This is understandable since the typical E\&M curriculum is dense with both fundamental insights, mathematical frameworks, and important technical applications. This structure leaves little space for a separate development of numerical recipes. 

Discretizing d'Alembert's method provides a means to combine the essential physics with FDTD in an insightful way. Constant velocity of waves in dispersionless dielectric materials, partial reflection and transmission at boundaries between materials with different indices of refraction, and reflection and attenuation through conducting materials are derived with no restriction to single frequency. No understanding of complex numbers, trigonometric functions, or exponential functions are required to understand these results. Placing D'Alembert's method on a grid  shows the physical effects of the finite difference approximation in a way that is easy to visualize. It also allows for easy derivation of some results that are more complicated with the traditional FDTD derivation, such as that FDTD is exact for dielectrics when the ratio of the spacial and temporal grid spacings is the wave speed,\cite{magicgridspacing} incorporating reflectionless boundary conditions,\cite{reflectionless} and showing the method retains second order accuracy when the grid spacing varies with position and the material parameters make sudden jumps across layer boundaries.

\subsection{Structure of this Paper}
We develop the basic physics of electromagnetic waves in matter and the FDTD method of solving Maxwell's in 1-D systems, through a series of steps, starting with homogeneous dispersionless dielectrics and ending with inhomogeneous conductors and dielectrics with dispersion. In Section \ref{Wave Pulses} we apply d'Alembert's method to electromagnetic waves in a homogeneous dispersionless dielectric, and show that wave pulses move at constant speed without changing their shape. In Section \ref{Inhomogeneous} we introduce inhomogeneity and apply the method to a system composed of layers of dispersionless dielectrics, with arbitrary changes in the index of refraction across layers. We show that such systems have exact solutions when the layers are of equal optical path length, as is the case for distributed Bragg reflectors. We also derive the formula for the partial reflection and transmission of wave pulses across boundaries between layers of differing material properties and layer lengths and derive the formulas for reflectionless boundary conditions at system boundaries. In Section \ref{Model}, we introduce conductivity and dispersion by deriving a model of wave pulses partially reflecting, transmitting, and attenuating through regions of constant current. We demonstrate the model for the case of a simple conductor and develop a succinct formula for the partial reflection, transmission, and attenuation of waves pulses through a conductor. In Section \ref{=2FDTD} we show that our method is equivalent to FDTD and that, despite appearances, it is second order accurate for grid spacings arbitrarily varying with position and for arbitrary jumps in material parameters across layer boundaries.

\section{Wave Pulse Approach for Solving Maxwell's Equations in One Dimensional Systems}\label{Wave Pulses}
d'Alembert's method presents a simple solution to the one-dimensional wave equation usually encountered in a first course on partial differential equations or the physics of waves. d'Alembert solved the wave equation by transforming to variables representing wave pulses that move forward and backwards through the media. In this section, using d'Alembert's method, we rewrite Maxwell's equations for electromagnetic waves in media in terms of wave pulses and solve them for a dispersionless dielectric. 
Maxwell's equations for electromagnetic waves in matter are
\begin{equation}
\mathbf{\nabla} \times \mathbf{E}= -\partial_t \mathbf{B},
\end{equation}
\begin{equation}
\frac{1}{\mu}\mathbf{\nabla} \times \mathbf{B}= \epsilon\partial_t \mathbf{E} +\sigma \mathbf{E} + \mathbf{J_p},
\end{equation}
where $\mathbf{E}$ and $\mathbf{B}$ are the electric and magnetic fields respectively. The material parameters $\mu$, $\epsilon$, and $\sigma$ are the magnetic permeability, electric permitivity, and electrical conductivity respectively. $\epsilon\partial_t \mathbf{E}$, incorporates the displacement current and polarization current. $\sigma\mathbf{E}$ is the conduction current. $\mathbf{J_p}$ is the remaining part of the current that includes any more complicated dependencies on the electric field, usually through a differential equation. In the simplest model of a dielectric material $\sigma=0$ and  $\mathbf{J_p} =0$. For the simplest model of a conductor $\mathbf{J_p} =0$. 

We consider only one dimensional systems in this paper, by which we mean that the media is homogeneous and infinite and all fields are non-varying in two of the spatial dimensions but can vary in the remaining spatial dimension. This allows us to consider all of the physics of wave pulses interacting with matter at normal incidence. Also, for simplicity, only constant $\mu$ is considered and only waves of one polarization are considered. For electromagnetic waves polarized in the y direction and moving in the x direction through materials that have constant $\mu$ and with other material properties varying only in the x direction, Maxwell's equations simplify to
\begin{equation}\label{m1}
\partial_x E_y(x,t)= -\partial_t B_z(x,t),
\end{equation}
\begin{equation}\label{m2}
\partial_xB_z(x,t)= -\mu\epsilon(x)\partial_t E_y(x,t) -\mu\sigma(x) E_y(x,t) - \mu J_{py}(x,t).
\end{equation}
In what follows we will suppress the dimensional indices and $x$ and $t$ variables on the field and current components. 

The variables that one chooses to represent physical phenomena can make some physics clear and other physics obscure. Maxwell's equations in terms of $E$ and $B$ fields clearly show that a changing magnetic field creates an electric field and a changing electric field creates a magnetic field. However, it is not immediately obvious by looking at these equations that wave pulses move through space with constant velocity. But we can see that explicitly if we change to d'Alembert's wave pulse variables
\begin{eqnarray}
F = \frac{E + vB}{2},\\
G = \frac{E - vB}{2},\\
E = F + G,\\
vB = F - G.
\end{eqnarray}
where the physical meaning of $v$,which has units of velocity, will be determined momentarily.
Taking the partial derivatives of $F$ and $G$ we arrive at
\begin{eqnarray}
\partial_t F + v\partial_x F = -\frac{\mu v^2}{2} J,\\
\partial_t G - v\partial_x G = -\frac{\mu v^2}{2}J,
\end{eqnarray}
where
\begin{equation}\label{Thecurrent}
J = (\epsilon -\frac{1}{\mu v^2})\partial_t E  +\sigma E + J_p.
\end{equation}

Using the directional derivatives,
\begin{eqnarray}
\partial_R &=& \partial_t +v\partial_x,\\
\partial_L &=& \partial_t -v\partial_x,
\end{eqnarray} 
in which we can think of $\partial_R$ as a forward change in time and a spacial change to the right and $\partial_L$ as a forward change in time and a spacial change to the left, Maxwell's equations for waves in matter become
\begin{eqnarray}
\partial_R F &=& -\frac{\mu v^2}{2}J,\\
\partial_L G &=& -\frac{\mu v^2}{2}J.
\end{eqnarray}
If we know the values of the $F$ and $G$ fields at a given position and time, we can determine the values at a different position and time by integrating the equations along their paths. We integrate the equation for F forward through time and to the right.  We integrate the equation for G forward through time and to the left.
\begin{equation}\label{exactf}
F(x_i+v(t_f-t_i),t_f) = F(x_i,t_i) - \frac{\mu v^2}{2}\int^{t_f}_{t_i} d\tau J(x_i+ v(\tau-t_i),\tau),
\end{equation}
\begin{equation}\label{exactg}
G(x_i-v(t_f-t_i),t_f) = G(x_i,t_i) - \frac{\mu v^2}{2}\int^{t_f}_{t_i} d\tau J(x_i- v(\tau-t_i),\tau).
\end{equation}
The $F$ wave pulse moves to the right with speed $v$ and the $G$ wave pulse moves to the left with speed $v$. The wave pulses $F$ and $G$ interact through the current $J$. For the special case of a simple dielectric with speed $v$ chosen so that $\epsilon\mu v^2=1$ then $J=0$ and the wave pulse equations simplify to
\begin{equation}
F(x_i+v(t_f-t_i),t_f) = F(x_i,t_i),
\end{equation}
\begin{equation}
G(x_i-v(t_f-t_i),t_f) = G(x_i,t_i).
\end{equation}
We choose grid spacing, $\Delta x = v \Delta t$. Using unitless integer spatial and temporal grid coordinates $X$ and $T$  and placing the spacial coordinate as a subscript and the temporal coordinate as a superscript so that $x_X = X v\Delta t$ and $t
^T= T\Delta t$, we have
\begin{figure}[h!]
\centering
\includegraphics{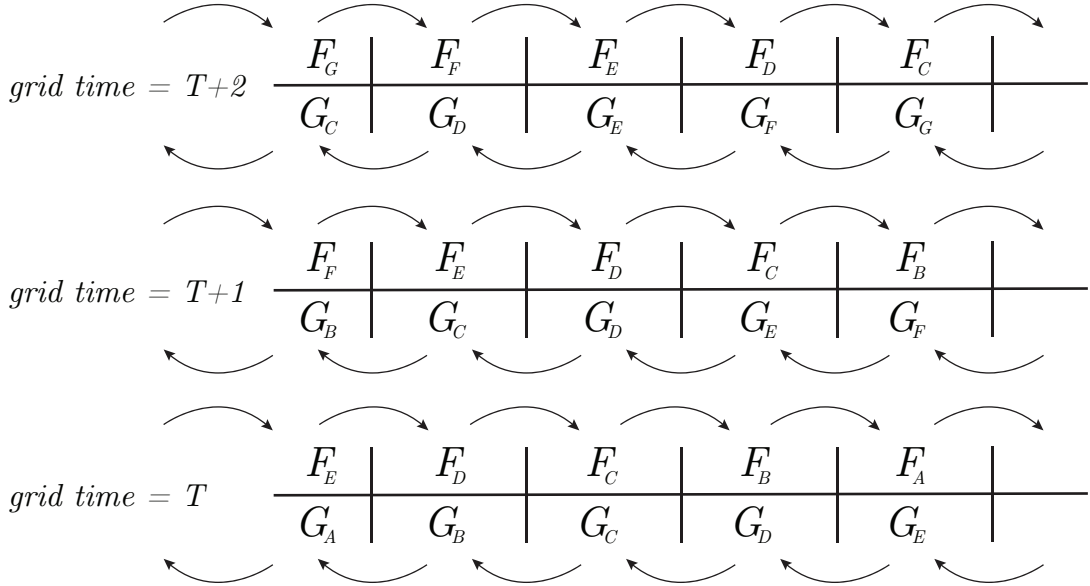}
\caption{The F wave pulses move to the right and the G wave pulses move to the left at constant speed through a simple dielectric material without changing their shape. Electric and magnetic fields are determined by adding and subtracting $F$ and $G$ wave pulses. For example, the electric field in the second position at grid time $T+1$ is $F_E+G_C$ and the magnetic field is $(F_E - G_C)/v$.}
\label{FigureConstantV}
\end{figure}
\begin{equation}
F^{T+1}_{X+1} = F^T_X,
\end{equation}
\begin{equation}
G^{T+1}_{X-1}= G^T_X.
\end{equation}
These equations look like a finite difference approximation but they are exact, and, as shown in Fig. \ref{FigureConstantV} completely express the physics that the $F$ wave pulses move to the right and the $G$ wave pulses move to the left at constant speed $v= (\mu\epsilon)^{-1/2}$ through a simple dielectric material without changing their shape. If a different grid spacing was chosen so that $\Delta x \neq (\mu\epsilon)^{-1/2}\Delta t$ then the current would not be zero, the update equations would not be exact, and there would be anomalous dispersion from the mismatch of the grid with the physical wave velocity. We, therefore, call the first term in Eq. (\ref{Thecurrent}), the grid current.

Reflectionless boundary conditions with an incoming rightward moving pulse are easily implemented in this formalism. The condition at the left end at position $X=1$ is satisfied by $F^T_1 = P^T$, where $P^T$ represents the amplitude of the incoming pulse at time $T$, and the condition at the right end at position $X = N$ is satisfied by $G^T_N=0$.

In the next section we apply d'Alembert's solution to Maxwell's equations to determine the fields in a system of layered dispersionless diectrics. In the process we derive the the formula for partial reflection and transmission at boundaries between materials with differing indices of refraction. 
\section{Layered Dispersionless Diectrics}\label{Inhomogeneous}
\subsection{Transport Through Layers of the Same Optical Path Length}
\begin{figure}[h!]
\centering
\includegraphics{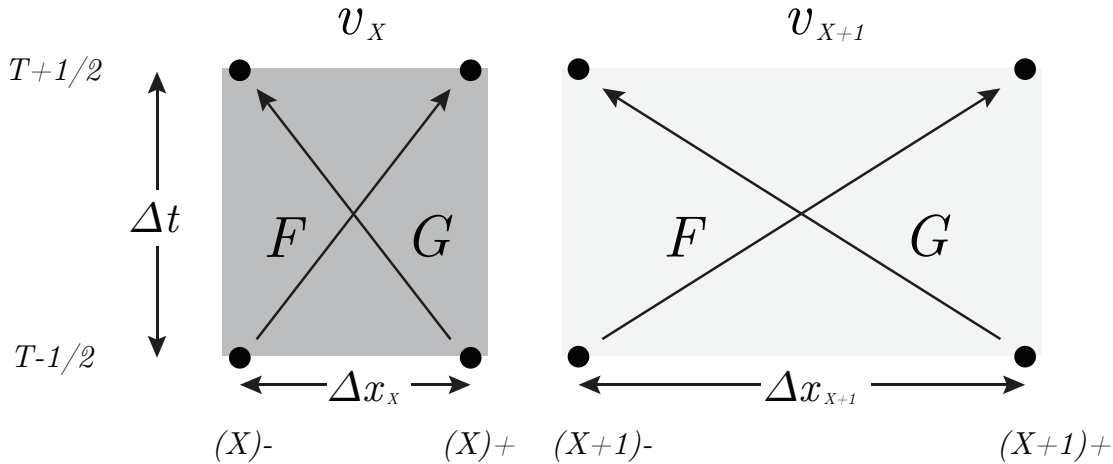}
\caption{For systems where each layer has a different velocity and thickness but the same optical path length, $\Delta x_X = v_X\Delta t$, $\Delta x_{X+1} =v_{X+1}\Delta t$, each F pulse moves across one layer to the right and each G pulse moves across one layer to the left in time $\Delta t$ without changing amplitude. We refer to the right end of layer $X$ as $(X)+$ and the left end of layer $X$ as $(X)-$.}
\label{FigureVaryingV}
\end{figure}
In the previous section, we determined results for a homogeneous dispersionless dielectric, meaning a dielectric with a permitivity independent of position and with $\sigma =0$ and $J_p=0$.  In this section we determine results for an inhomogeneous dispersionless dielectric, meaning a dielectric where the permitivity is a function of the $x$ coordinates. Our solution is exact provided that three conditions are satisfied.  The first is that we continue dealing with dispersionless dielectrics, $\sigma =0$ and $J_p=0$. The second is that the grid spacing is small enough that the permitivity is approximately constant throughout each grid spacing. This is a common numerical approximation. We call the space between grid points `layers' and refer to the constant permitivity in the Xth layer as $\epsilon_X$ where the integer grid position $X$ occurs at the center of the $X$th layer. Our third condition is less common. It is that the grid currents equal zero, meaning that the Xth layer has length $\Delta x_X = v_X \Delta t$, where $\epsilon_X\mu v_X^2 =1$. This means that the time it takes for light to travel though each layer is the same for each layer even if their permitivities, and therefore their speeds of light, differ. Dielectric layers with a faster speed of light will be longer and dielectric layers with a slower speed of light will be shorter so that light takes the same time, $\Delta t$, to travel through each layer. In other words, the optical path lengths are the same. There are important layered systems, such as Bragg reflectors,\cite{Heavens,Griffiths2,Girvin,Horsley} where this condition is satisfied, and which motivated the development of this method. For layers with equal optical path lengths, with the convention that unitless integer grid time coordinates, $T$, occur when pulses are at layer centers, pulses will be at layer boundaries at unitless grid times $T \pm 1/2$. We refer to the right end of layer $X$ as $(X)+$ and the left end of layer $X$ as $(X)-$. As before, $t^T = T\Delta t$. But now, $x_{(X)+} = x_{X} + v_X\Delta t/2$ and $x_X = x_{(X)-} + v_X\Delta t/2$. As shown in Fig. \ref{FigureVaryingV}, for systems where each layer has the same optical path length, after a time $\Delta t$ each $F$ pulse moves across one layer to the right and each $G$ pulse moves across one layer to the left without changing their amplitudes. The equations that describe this transport of light across simple dielectric layers are
\begin{equation}\label{updatefsharp}
F^{T+1/2}_{(X)+} = F^{T-1/2}_{(X)-},
\end{equation}
\begin{equation}\label{updategflat}
G^{T+1/2}_{(X)-} = G^{T-1/2}_{(X)+}.
\end{equation}
For a Bragg reflector, each layer spacing in the simulation can equal the entirety of an actual material layer.

We next derive what happens to a pulse when it crosses from one layer to another. 
\subsection{Reflection from and Transmission through Layer Boundaries} 
\begin{figure}[h!]
\centering
\includegraphics{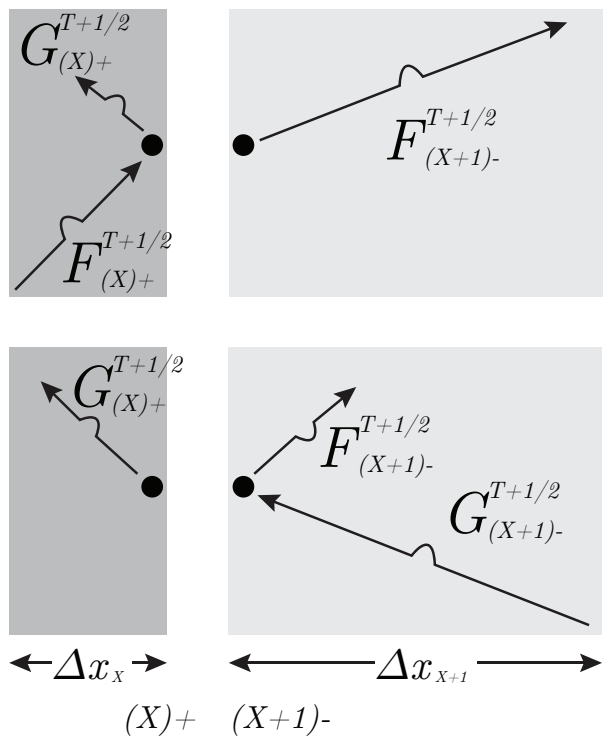}
\caption{Wave pulses partially reflect and partially transmit through layer boundaries. In the upper (lower) figure, a pulse that transmits through a boundary into a layer with a lower (higher) index of refraction is amplified (attenuated). The reflected pulse undergoes a 180 degree phase change if the layer the pulse reflects off of has a higher index of refraction, as is the case in the lower figure.}
\label{FigureBoundaries}
\end{figure}To determine how $F$ and $G$ wave pulses partially reflect from and partially transmit through the boundaries between layers, we integrate Maxwell's Eqs. (\ref{m1}),(\ref{m2}) across the $(X+1/2)$ layer boundary at time $T+1/2$ producing
\begin{equation}
E^{T+1/2}_{(X+1)-} - E^{T+1/2}_{(X)+} = -\int^{x_{(X+1)-}}_{x_{(X)+}}dx'\partial_t B(x',t^{T+1/2}),
\end{equation}
\begin{equation}\label{bboundary}
B^{T+1/2}_{(X+1)-} - B^{T+1/2}_{(X)+} = -\int^{x_{(X+1)-}}_{x_{(X)+}}dx'\left[\mu\epsilon(x')\partial_t E(x',t^{T+1/2}) +\mu\sigma(x') E(x',t^{T+1/2}) + \mu J_{p}(x',t^{T+1/2})\right].
\end{equation}
Both $(X)+$ and $(X+1)-$ correspond to $(X+1/2)$, the same position. So the integration intervals are zero. This makes the integrals zero, guaranteeing that $E$ and $B$ fields are continuous regardless if the material parameters are continuous or discontinuous across boundaries.\cite{surfacecurrents} The continuity of the electric field across the $(X+1/2)$ layer boundary produces
\begin{equation}
\label{boundaryE}
 E^{T+1/2}_{(X+1/2)} = F^{T+1/2}_{(X)+} + G^{T+1/2}_{(X)+} = F^{T+1/2}_{(X+1)-} + G^{T+1/2}_{(X+1)-}.
\end{equation}
Defining $n_X= \frac{c}{v_X}$, which is the index of refraction for layer $X$, the continuity of the magnetic field across the same boundary produces
\begin{equation}\label{boundaryB}
cB^{T+1/2}_{(X+1/2)} = n_X( F^{T+1/2}_{(X)+} - G^{T+1/2}_{(X)+})= n_{X+1}(F^{T+1/2}_{(X+1)-} - G^{T+1/2}_{(X+1)-}).
\end{equation}
Combining these so that wave pulses coming out of a boundary are determined by the wave pulses going into a boundary we have
\begin{equation}
\label{updatefflat}
F^{T+1/2}_{(X+1)-} = \frac{2n_X}{n_X+n_{X+1}} F^{T+1/2}_{(X)+} + \frac{n_{X+1}-n_X}{n_X+n_{X+1}}G^{T+1/2}_{(X+1)-},
\end{equation}
\begin{equation}
\label{updategsharp}
G^{T+1/2}_{(X)+} = \frac{2n_{X+1}}{n_X+n_{X+1}} G^{T+1/2}_{(X+1)-} + \frac{n_X-n_{X+1}}{n_X+n_{X+1}}F^{T+1/2}_{(X)+}.  
\end{equation}
which captures all of the physics of transmission and reflection at boundaries between dielectric layers of different indices of refraction.  For example, consider the situation where there is an incoming pulse to the right, $F^{T+1/2}_{(X)+}\neq 0$ and no incoming pulse to the left, $G^{T+1/2}_{(X+1)-}=0$. Equation (\ref{updatefflat}) shows that a pulse that transmits through a boundary into a layer with a lower (higher) index of refraction is amplified (attenuated). Equation (\ref{updategsharp}) shows that some of a pulse is reflected and the reflected pulse undergoes a 180 degree phase change if the layer the pulse reflects off of has a higher index of refraction. As the difference in the two indices increases the proportion of the wave that is reflected increases. As the differences of the two indices approach zero, the wave pulse is transmitted unchanged and without any reflection. A visualization is shown in Fig. \ref{FigureBoundaries}.

Reflectionless boundary conditions at both ends with an incoming pulse entering on the leftmost end are satisfied with 
\begin{equation}\label{reflectionlessF1}
F^{T+1/2}_{(1)-} = P^{T+1/2},
\end{equation}
\begin{equation}\label{reflectionlessGN}
G^{T+1/2}_{(N)+}=0.
\end{equation} 

\section{The Constant Current Model and Absorption and Partial Reflection through a Conductor}\label{Model}
\subsection{The Constant Current Model Inside Layers}
\begin{figure}[h!]
\centering
\includegraphics{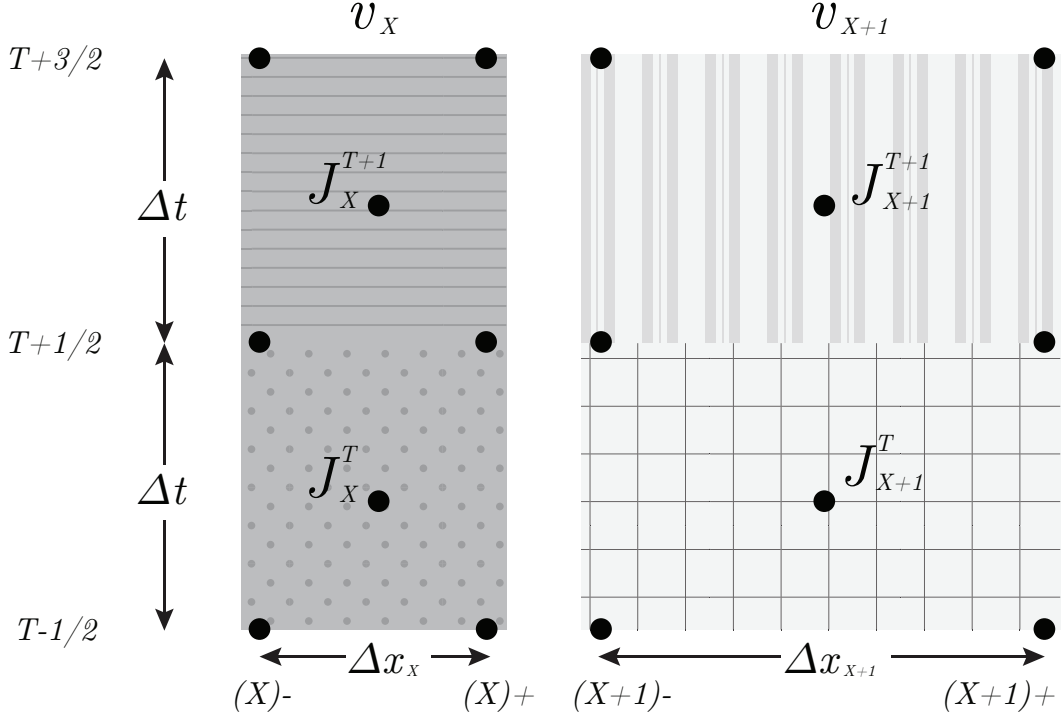}
\caption{In the piece-wise constant current model, the current is constant within each layer but changes its value in time intervals of $\Delta t$, as represented by the different patterns in the figure.}
\label{FigureConstantJ}
\end{figure}
When the system being simulated includes conducting layers, or dielectrics with dispersion, or it is not possible to set the grid spacing so that the grid current is zero, the update Eqs. (\ref{updatefsharp}),(\ref{updategflat}) will no longer be correct.  We present a piece-wise constant current model for such systems, in which, as seen in Fig. \ref{FigureConstantJ}, the current is constant within each layer but changes in time intervals of $\Delta t$. This is a reasonable approximation provided the layer lengths and $\Delta t$ are small. We will later show that the model is equivalent to FDTD. 

We return to Eqs. (\ref{exactf}),(\ref{exactg}), for materials for which the current $J$ is not zero. Using our layer notation,
\begin{equation}\label{integralf}
F^{T+1/2}_{(X)+} = F^{T-1/2}_{(X)-} - \frac{\mu v_X^2}{2}\int^{t^{T+1/2}}_{t^{T-1/2}} d\tau J(x+ v(\tau-t),\tau),
\end{equation}
\begin{equation}\label{integralg}
G^{T+1/2}_{(X)-} = G^{T-1/2}_{(X)+} - \frac{\mu v_X^2}{2}\int^{t^{T+1/2}}_{t^{T-1/2}} d\tau J(x- v(\tau-t),\tau).
\end{equation}
If the current is constant and equal to $J^T_X$ everywhere in layer $X$ from $(X)-$ to $(X)+$ and in time from $T-1/2$ to $T+1/2$ then the above equations are easily integrated. 
The wave pulses at one end of a boundary depend on the other end via,
\begin{equation}
\label{updatefsharp2}
F^{T+1/2}_{(X)+} = F^{T-1/2}_{(X)-} - \frac{\mu v_X^2\Delta t }{2}J^T_X,  
\end{equation}
\begin{equation}
\label{updategflat2}
G^{T+1/2}_{(X)-} = G^{T-1/2}_{(X)+}  -  \frac{\mu v_X^2\Delta t }{2}J^T_X.
\end{equation}
The above equations require us to know the current at the center of each layer. The current at the center of each layer is dependent on the electric field at the center of the layer, which we find by integrating the $F$ and $G$ equation from the layer boundaries at time $T -1/2$ to their centers at time $T$,
\begin{equation}
F^T_X = F^{T-1/2}_{(X)-} - \frac{\mu v_X^2\Delta t }{4}J^T_X,
\end{equation}
\begin{equation}
G^T_X = G^{T-1/2}_{(X)+} - \frac{\mu v_X^2\Delta t }{4}J^T_X.
\end{equation}

Adding the equations together produces
\begin{equation}\label{updateec}
E^T_X = F^{T-1/2}_{(X)-} + G^{T-1/2}_{(X)+} - \frac{\mu v_X^2\Delta t }{2}J^T_X.
\end{equation}

It will be convenient to express the wave pulses at the boundaries in terms of the electric field at the center instead of the current. This can be done by replacing the current term in Eq. (\ref{updatefsharp2}),(\ref{updategflat2}) with its dependence on $E^T_X$ from Eq.(\ref{updateec}).  The resulting update equations for $F$ and $G$ at the boundaries are
\begin{equation}\label{nojf}
F^{T+1/2}_{(X)+} = E^T_X  - G^{T-1/2}_{(X)+},
\end{equation}
\begin{equation}\label{nojg}
G^{T+1/2}_{(X+1)-} = E^T_{X+1}  - F^{T-1/2}_{(X+1)-}.
\end{equation}
\subsection{The Constant Current Model at Layer and System Boundaries}
Equations (\ref{updatefflat}),(\ref{updategsharp}), for wave pulse reflection and transmission at layer boundaries, holds in the constant current model for pulses of any shape and for boundaries between any two materials, including conductors and dielectrics with dispersion because the only physics used to derive them is the continuity of the electric and magnetic fields across boundaries.\cite{alternatives} In cases where the materials are not dispersionless dielectrics, or where the grid current is not zero, the index of refraction, $n_X= \frac{c}{v_X} = \frac{c\Delta t}{\Delta x_X}$, in Eqs.(\ref{updatefflat}),(\ref{updategsharp}) is a property of the grid, not necessarily related to the phase or group wave velocity in any way. 

Reflectionless boundary conditions for pulses at positions $X=1$ and $X=N$ are determined from Eqs.(\ref{nojf}),(\ref{nojg}) together with Eqs.(\ref{reflectionlessF1}),(\ref{reflectionlessGN}),
\begin{eqnarray}
F^{T+1/2}_{(1)-} &=& P^{T+1/2},\label{rlessF1}\\
G^{T+1/2}_{(1)-} &=& E^T_1  - P^{T-1/2},\label{rlessG1}\\
F^{T+1/2}_{(N)+} &=& E^T_N,\label{rlessFN}\\
G^{T+1/2}_{(N)+} &=& 0.\label{rlessGN}
\end{eqnarray}
Equations (\ref{updatefflat}),(\ref{updategsharp}),(\ref{updateec})--(\ref{rlessGN}), form a complete set of equations for the wave pulses in the constant-current model.  

To proceed further we produce explicit formulas for the constant currents $J^T_X$.  
\subsection{Absorption and Partial Reflection through a Conductor}\label{Conductor Example}
\begin{figure}[h!]
\centering
\includegraphics{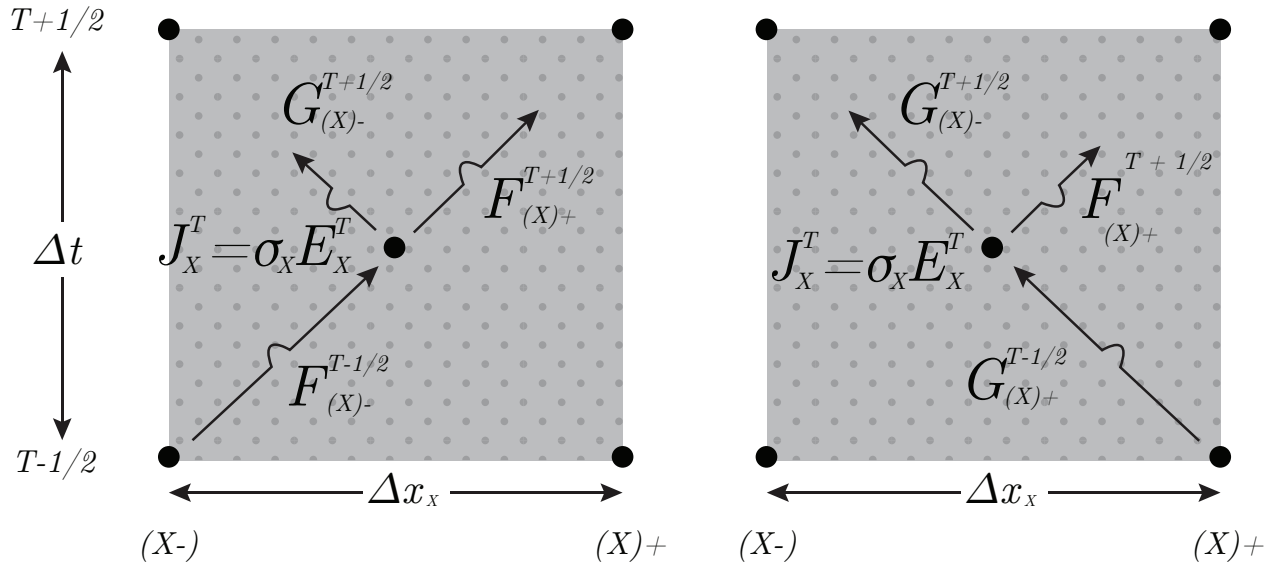}
\caption{Wave pulses partially reflect and partially transmit through a conducting layer. A pulse that transmits through a conducting layer is always attenuated. The reflected pulse always undergoes a 180 degree phase change. Comparison with Fig. \ref{FigureBoundaries} demonstrates that a wave pulse moving through a conducting layer acts like a pulse in a dielectric layer transmitting through and reflecting from a boundary to a dielectric layer of higher index of refraction.}
\label{FigureConductor}
\end{figure}We now show how the constant current model illuminates the physics of wave pulses moving through a conductor. Choosing layer lengths such that the grid current is zero, $\epsilon_X\mu v_X^2 =1$, and considering a simple conductor, $J_p$ =0, the current, $J^T_X$ from Eq.(\ref{Thecurrent}), reduces to 
\begin{equation}
J^T_X = \sigma_X E^T_X.
\end{equation}

Plugging this $J^T_X$ into Eq.(\ref{updateec}), solving for $E^T_X$, and
using it in the equation for the wave pulses at the boundaries, Eqs. (\ref{nojf}),(\ref{nojg}), we have 
 \begin{equation}\label{conduct1}
 F^{T+1/2}_{(X)+}= \frac{1}{1+ \alpha_X}F^{T-1/2}_{(X)-} - \frac{\alpha_X}{1+ \alpha_X} G^{T-1/2}_{(X)+},
\end{equation}
\begin{equation}\label{conduct2}
   G^{T+1/2}_{(X)-} = \frac{1}{1+ \alpha_X}G^{T-1/2}_{(X)+}  - \frac{\alpha_X}{1+ \alpha_X}F^{T-1/2}_{(X)-},
\end{equation}
where
\begin{equation}
\alpha_X = \frac{\mu v^2_X\sigma_X\Delta t}{2} = \frac{\sigma_X\Delta t}{2\epsilon_X}.
\end{equation}
These equations show all of the qualitative effects of conductivity. For example, consider the situation where there is an incoming pulse to the right, $F^{T-1/2}_{(X)-}\neq 0$, and no incoming pulse to the left, $G^{T-1/2}_{(X)+}=0$. Equation (\ref{conduct1}) shows that a pulse that transmits through a conducting layer is always attenuated. Equation (\ref{conduct2}) shows that the the reflected pulse always undergoes a 180 degree phase change. As the conductivity goes to infinity, all of the pulse is reflected and no transmission occurs, which is why metals make good mirrors. As the conductivity goes to zero, the wave pulse is transmitted without any attenuation or reflection. Comparison with Eqs.(\ref{updatefflat}),(\ref{updategsharp}) demonstrates that a pulse in a conducting layer acts like a pulse at a boundary between dielectric layers going from the lower to the higher index dielectric, with $\alpha = (n_H -n_L)/(2n_L)$. A visualization is shown in Fig. \ref{FigureConductor}.

In the next section we extend these results for more general currents of the form Eq.(\ref{Thecurrent}), including when the grid current is not zero. In the process we demonstrate equivalence with FDTD.

\section{Equivalence of the Constant Current Model and FDTD}\label{=2FDTD}
\subsection{The Electric Field at Layer Centers}
We incorporate constant currents that depend on the time derivative of the electric field, which occurs when the grid current and/or $J_p$ is not zero, by integrating in time the formula for $J$,  Eq.(\ref{Thecurrent}), from $T-1$ to $T$ at layer center $X$,
\begin{equation}
\int^{t^T}_{t^{T-1}}d\tau J(x,\tau) = (\epsilon_X -\frac{1}{\mu v_X^2})\int^{t^T}_{t^{T-1}}d\tau\partial_\tau E(x,\tau)  +\sigma_X\int^{t^T}_{t^{T-1}}d\tau E(x,\tau) +\int^{t^T}_{t^{T-1}}d\tau J_p(x,\tau).
\end{equation}
The first term on the RHS is integrated exactly. All other terms are evaluated via the constant current model, or equivalently, the average end point approximation of the integral, which is accurate to second order in $\Delta t$,
\begin{equation}\label{currentequation}
\Delta t\frac{J^T_X + J^{T-1}_X}{2}= (\epsilon_X - \frac{1}{\mu v_X^2})(E^T_X - E^{T-1}_X) +\Delta t\sigma_X \frac{E^T_X+ E^{T-1}_X}{2} + \Delta t\frac{J^T_{px} + J^{T-1}_{px}}{2} + O(\Delta t\,^3).
\end{equation}
This expression for $J^T_X$ in terms of $J^{T-1}_X$ can be used in Eq.(\ref{updateec}), which we show in subsection \ref{2ndOrder} is also accurate to second order for nonzero $\sigma$ and $J_p$.

Alternatively, we can remove any explicit dependence of $J$ in the past, so that we do not need to keep it in memory as a separate variable. To do this, consider the electric field at layer center $X$ at time $T$ and transport its component wave pulses backwards in time to reproduce Eq.(\ref{updateec})
\begin{equation}\label{newupdateec}
E^T_X = F^{T-1/2}_{(X)-} + G^{T-1/2}_{(X)+} - \frac{\mu v_X^2\Delta t }{2}J^T_X,
\end{equation}
Following a derivation similar to that for Eq.(\ref{newupdateec}), consider the electric field at layer center $X$ at time $T -1$ and transport its component wave pulses forward in time,
\begin{equation}
E^{T-1}_X = F^{T-1/2}_{(X)+} + G^{T-1/2}_{(X)-} + \frac{\mu v_X^2\Delta t }{2}J^{T-1}_X,
\end{equation}
Subtracting fields and rewriting the RHS's in terms of E and B fields we have
\begin{equation}\label{secondorderE}
E^T_X- E^{T-1}_X = v_X\left(B^{T-1/2}_{(X)-} - B^{T-1/2}_{(X)+}\right) - \frac{\mu v_X^2\Delta t }{2}(J^T_X+J^{T-1}_X).
\end{equation}

Placing the expression for the current, Eq.(\ref{currentequation}), into the expression for the difference of the electric fields, Eq.(\ref{secondorderE}), and dividing by $v_X^2 = (\frac{\Delta x_X}{\Delta t})^2$ produces
\begin{equation}\label{FDTD1}
\mu\epsilon_X(E^T_X - E^{T-1}_X) = - \Delta t\frac{B^{T-1/2}_{(X)+} - B^{T-1/2}_{(X)-}}{\Delta x_X} -\Delta t\mu\sigma_X \frac{E^T_X+ E^{T-1}_X}{2} - \Delta t\mu\frac{J^T_{px} + J^{T-1}_{px}}{2},
\end{equation}
which is the standard FDTD update formula for the electric field at layer centers.\cite{Yee, Taflove, Sipos, Visscher, Gould, Kunz} 

\subsection{The Magnetic Field at Layer Boundaries}
We derive the update equation for the magnetic field at layer boundaries and system boundaries by combining the equations for wave pulse propagation within a layer, Eqs.(\ref{nojf}),(\ref{nojg}), with the equations of continuity of electric and magnetic fields across boundaries, Eqs.(\ref{boundaryE}),(\ref{boundaryB}), keeping in mind that the wave velocity on either side of a boundary may be different. 

Subtracting Eq.(\ref{nojf}) for layer $X$ and Eq.(\ref{nojg}) for layer $X+1$ produces, 
\begin{equation}\label{subtract}
F^{T+1/2}_{(X)+} - G^{T+1/2}_{(X+1)-} = E^T_X -E^T_{X+1}  - G^{T-1/2}_{(X)+} + F^{T-1/2}_{(X+1)-},
\end{equation}
Using the relationship between fields and wave pulses at boundaries, Eqs.(\ref{boundaryE}),(\ref{boundaryB}), the magnetic field at time $T+1/2$ at layer boundary $(X+1/2)$ in terms of wave pulses entering the boundary is
\begin{equation}
\frac{v_X + v_{X+1}}{2}B^{T+1/2}_{(X+1/2)} = F^{T+1/2}_{(X)+} -G^{T+1/2}_{(X+1)-}
\end{equation}
and the magnetic field at time $T-1/2$ at layer boundary $(X+1/2)$ in terms of wave pulses leaving the boundary is
\begin{equation}
\frac{v_X + v_{X+1}}{2}B^{T-1/2}_{(X+1/2)} = F^{T-1/2}_{(X+1)-}-G_{(X)+}^{T-1/2}.
\end{equation}
Replacing wave pulse differences in Eq.(\ref{subtract}) with $B$ fields, and dividing by $1/2\left(v_X +v_{X+1}\right) = 1/2\left(\frac{\Delta x_X}{\Delta t} +\frac{\Delta x_{X+1}}{\Delta t}\right)$ produces the simplest generalization of the traditional FDTD equation for the magnetic field\cite{Yee, Taflove, Sipos, Visscher, Gould, Kunz} between layers with variable grid spacing,
\begin{equation}\label{FDTD2}
B^{T+1/2}_{(X)+} -  B^{T-1/2}_{(X)+} = -\Delta t\frac {E^T_{X+1}  - E^T_X}{ 1/2\left(\Delta x_X +\Delta x_{X+1}\right)}.
\end{equation}

Reflectionless boundary conditions for the magnetic fields at the ends of the system are determined from Eqs.(\ref{rlessF1})--(\ref{rlessGN}), 
\begin{equation}\label{reflectionless1}
v_1B^{T+1/2}_{(1/2)} = F^{T+1/2}_{(1)-} -G^{T+1/2}_{(1)-} = P^{T+1/2} - E^T_1  + P^{T-1/2},
\end{equation}
\begin{equation}\label{reflectionlessN}
v_NB^{T+1/2}_{(N+1/2)} = F^{T+1/2}_{(N)+} - G^{T+1/2}_{(N)+} = E^T_N.
\end{equation}
The results of this section show that the traditional FDTD method is equivalent to the d'Alembert's method with the piece-wise constant current. This provides a physical model of FDTD inexactness. Everything that is true about one method is true about the other. In particular, the FDTD method is exact for dispersionless dielectrics where each layer is of equal optical path length, and for a model system composed of conducting layers and dielectrics with dispersion with piece-wise constant current in each layer. 

Next we show that the update equations are accurate to second order when the currents are not constant. 
\subsection{Proof of Second Order Accuracy}\label{2ndOrder}
In this section we show that the constant current model, and equivalently, the traditional FDTD method with the simplest generalization to variable grid spacing, is second order accurate. This is not immediately obvious because approximating the current in Eq. (\ref{integralf}) by the first terms of its Taylor expansion, and integrating half way across the layer, produces
\begin{equation}\label{taylorf}
F^T_X = F^{T-1/2}_{(X)-} - \frac{\mu v_X^2\Delta t }{4}J^T_X + \frac{\mu v_X^2(\Delta t)^2}{8} \partial_R J^T_X + O(\Delta t\,^3).
\end{equation}
The equation has a second order term proportional to $\partial_R J^T_X$ that is not part of the constant current model and, if non-zero, makes the constant current model only first order accurate.  We now show that the offending term is actually third order, if it was zero at the initial time when the simulation began, by subtracting the same expression one time step in the past,
\begin{equation}\label{2to3Begin}
\partial_R J^T_X - \partial_R J^{T-1}_X  =  \int^{t^T}_{t^{T-1}}d \tau\partial_\tau \partial_R J(x_X,\tau) = \Delta t\partial_t \partial_RJ^{T-1/2}_X + O(\Delta t\,^3).
\end{equation}
The RHS was determined by presuming that, since $J$ is linearly dependent on $E$ and the material properties do not change in time, the integrand changes smoothly. Therefore, the second order accurate midpoint approximation to the integral can be used. We now have the result
\begin{equation}
\frac{\mu v_X^2(\Delta t)^2}{8} \partial_R J^T_X = \frac{\mu v_X^2(\Delta t)^2}{8} \partial_R J^{T-1}_X + O(\Delta t\,^3).
\end{equation}
The left and right sides are the same except for the time, indicating that each side is equal to the same quantity, which is constant in time, where `constant' in this context means differing by terms third order in $\Delta t$. If the initial conditions are that there are no fields and currents at the initial time, the RHS at time $T=0$ is zero. The constant will remain zero to second order accuracy for each time step, even when fields pass through the system.  Therefore we can dispense with the term proportional to $\partial_R J$ in Eq. (\ref{taylorf}), and the second order accurate update equation for $F$ is now what one would derive from the constant current model. The same procedure works with the equation for $G$ to dispense with the term proportional to $\partial_L J$.  

The result of this section is that the constant current model, and therefore the traditional FDTD method with the simplest generalization to variable grid spacing, is accurate to second order, even when the grid spacing varies with position and the material parameters make sudden jumps between layers.  The price we have paid for this is the insistence that the initial conditions for the system must be field-less and current-less, which is true in the system interior for most simulations.  

\section{Summary}
The Finite Difference Time Domain method for numerically simulating electromagnetic waves in matter for one dimensional systems was derived using d'Alembert's solution to the wave equation. The method utilizes forward and backward moving wave pulse variables instead of electric and magnetic field variables.   Constant velocity of waves in dispersionless dielectric materials,Eqs.(\ref{updatefsharp}),(\ref{updategflat}), partial reflection and transmission at boundaries between materials with different indices of refraction, Eqs.(\ref{updatefflat}),(\ref{updategsharp}), and reflection and attenuation through conducting materials, Eqs.(\ref{conduct1}),(\ref{conduct2}), were derived in the process of deriving the method so that real physics can be learned simultaneously with the numerics.   The traditional FDTD equations were derived from a piece-wise constant current model showing the physical effects of the finite difference approximation in a way that is easy to visualize. d'Alembert's method allowed for easy derivation of some results that are more complicated with finite differences alone such as showing that FDTD is exact for dielectrics when the ratio of the spacial and temporal grid spacings is the wave speed, Eqs.(\ref{updatefsharp}),(\ref{updategflat}), deriving reflectionless boundary conditions, Eqs.(\ref{rlessF1})--(\ref{rlessGN}),(\ref{reflectionless1}),(\ref{reflectionlessN}), and showing the method retains second order accuracy for grid spacing that varies with position and when the material parameters make sudden jumps across boundaries, Sec. \ref{2ndOrder}.

\begin{acknowledgments}
R.H and N.P.S. were supported by the National Science Foundation's MRSEC program (DMR-1720319) at the Materials Research Center at Northwestern University. R.H. was also supported by the Robert Noyce Teacher Scholarship program (DUE‐1439761) at the Department of Curriculum and Instruction at the University of Illinois at Chicago. The statements made and views expressed in this article are solely the responsibilities of the authors and do not reflect the views of the National Science Foundation.
\end{acknowledgments}


\begin{thebibliography}{99}
\bibitem{d'Alembert1}
Jean Le Rond D'Alembert, ``Recherches sur la courbe que forme une corde tendue mise en vibration (Researches on the curve that a tense cord forms when set into vibration),'' Histoire de l'academie royale des sciences et belles lettres de Berlin, \textbf{3}, 214--219 (1747). 

\bibitem{d'Alembert2}
Jean Le Rond D'Alembert, ``Suite des recherches sur la courbe que forme une corde tendue mise en vibration (Further researches on the curve that a tense cord forms when set into vibration),'' Histoire de l'academie royale des sciences et belles lettres de Berlin, \textbf{3}, 220--249 (1747).

\bibitem{d'Alembert3}
Jean Le Rond D'Alembert, ``Addition au memoire sur la courbe que forme une corde tendue mise en vibration,'' Histoire de l'academie royale des sciences et belles lettres de Berlin, \textbf{6}, 355--360 (1750).

\bibitem{Wheeler}
Gerald F. Wheeler and William P Crummett, ``The Vibrating String Controversy,'' Am. J. Phys. \textbf{55}, 33-37 (1987); doi: 10.1119/1.15311

\bibitem{Yee} 
Kane Yee, ``Numerical solution of initial boundary value problems involving Maxwell's equations in isotropic media,'' IEEE Transactions on Antennas and Propagation \textbf{14} (3), 302--307 (1966).

\bibitem{Taflove}
A. Taflove and S. C. Hagness, \textit{Computational Electrodynamics: The Finite-Difference Time-Domain Method}, 3rd  edition (Artech House, Norwood, MA, 2005). 

\bibitem{Sipos}
M. Sipos and B.G. Thompson, ``Electrodynamics on a grid: The finite-difference time-domain method applied to optics and cloaking,'' Am. J. Phys. \textbf{76}, 464-469 (2008); doi: 10.1119/1.2835056

\bibitem{Visscher}
P. B. Visscher, \textit{Fields and Electrodynamics: A Computer-Compatible Introduction}, (Wiley, New York, 1988). 

\bibitem{Gould}
H. Gould, J. Tobochnik, and W. Christian, \textit{Introduction to Computer Simulation Methods}, (Addison-Wesley, San Francisco, 2007), Chap. 10.

\bibitem{Kunz}
K. S. Kunz and R. J. Luebbers, \textit{The Finite Difference Time Domain Method for Electromagnetics}, (CRC Press, Boca Raton, FL, 1993). 

\bibitem{tightpulse}
R. M. Joseph, S. C. Hagness, and A. Taflove, ``Direct time integration of Maxwell's equations in linear dispersive media with absorption for scattering and propagation of femtosecond electromagnetic pulses,'' Optics  Letters \textbf{16}, 1412--1414, (Sept. 15, 1991).

\bibitem{atom}
S.-H. Chang and A. Taflove, ``Finite-difference time-domain model of lasing action in a four-level two-electron atomic system,'' Optics Express, \textbf{12} (16) 3827--3833, (Aug. 9,  2004).

\bibitem{Jackson}
J. D. Jackson, \textit{Classical electrodynamics}, (Wiley, New York, 1999) ISBN: 9780471309321

\bibitem{Griffiths}
D. J. Griffiths, \textit{Introduction to electrodynamics}, (Cambridge University Press, Cambridge, 2013). 

\bibitem{magicgridspacing} 
Taflove Ibid. Chapter 2, Section 2.5.

\bibitem{reflectionless}
Taflove Ibid. Chapter 6.

\bibitem{Heavens}
O. S. Heavens, \textit{Optical properties of thin solid films} 2nd edition, (Dover Publications, New York, 1991) ISBN 978-0486669243.

\bibitem{Griffiths2}
D. J. Griffiths and C. A. Steinke, ``Waves in locally periodic media,'' Am. J. Phys. \textbf{69}, 137--154 (2001). https://doi.org/10.1119/1.1308266

\bibitem{Girvin}
S. Girvin, and K. Yang, \textit{Modern Condensed Matter Physics}, (Cambridge University Press, Cambridge, 2019), pg 158. doi:10.1017/9781316480649.008

\bibitem{Horsley}
S. A. R. Horsley, J.-H. Wu, M. Artoni and G. C. La Rocca, ``Revisiting the Bragg Reflector to Illustrate Modern Developments in Optics,''
Am. J. Phys \textbf{82}, 206-213 (2014); https://doi.org/10.1119/1.4832436

\bibitem{surfacecurrents}
The argument is not true if a surface current exists in the infinitesimally small space between layers via a delta function multiplying a surface current term in the integrand. Surface currents do not arise at the boundaries between dispersionless dielectrics. For other materials, all currents are modeled inside layers.

\bibitem{alternatives} 
Treating surface currents inside layers requires that spacings be made sufficiently small that surface currents can be accurately modeled. Alternative methods (Taflove Ibid. Chapter 10) treat surface currents with a delta function at layer boundaries so that the grid spacings can be set to a much larger size. For these alternatives, the integral in Eq.(\ref{bboundary}) will not be zero, leading to a modification of Eqn's.(\ref{updatefflat},\ref{updategsharp}) and of FDTD not covered in this paper.

\end{thebibliography}
\end{document}